\documentclass[aps,prb,reprint,showpacs,groupedaddress]{revtex4-1}

\usepackage{dcolumn}
\usepackage{amsmath}
\usepackage{amssymb}
\usepackage{graphicx}
\usepackage{color}

\begin{document}

\title{Magneto-optical properties of bilayer transition metal dichalcogenides}
\author{M. Zubair$^{1}$, M. Tahir$^{2}$, and P. Vasilopoulos$^{1}$ }
\affiliation{$^{1}$Department of Physics, Concordia University, 7141 Sherbrooke Ouest, Montreal, Quebec H4B 1R6, Canada}
\affiliation{$^{2}$Department of Physics, Colorado State University, Fort Collins, CO 80523, USA}

\begin{abstract}

In    transition metal dichalcogenides   the spin-orbit interaction   affects differently the conduction  and valence band energies  as  functions of $k$ and the band gap is large. Consequently, when a perpendicular magnetic field $B$ is applied the conduction and valence band Landau levels  are also different and this leads to a splitting of the interband optical absorption lines 
in both the absence and presence of an external electric field $E_{z}$. When $B$ and $E_{z}$ are present 
the peaks in the imaginary part of the Hall conductivity give two distinct contributions of opposite sign to the interband spectrum. The real part of the right- and left-handed interband conductivity, however, retains its two-peak structure but the peaks are shifted in energy and amplitude with respect to each other in contrast with graphene. The response of the intraband conductivity is significantly modified when the Fermi energy $E_{F}$ and the  field $B$ are varied. Its optical spectral weight is found to increase with  $E_{F}$ in contrast with the decrease observed in  graphene. Further, 
the position and amplitude of the intraband response depends on the  field $B$. The absorption peaks 
vary linearly with $B$ for all fields 
 similar to bilayer graphene for low fields but in contrast  with the high-field  $\sqrt{B}$ dependence in it. 

\end{abstract}

\maketitle

\section{introduction}

Two-dimensional materials have  attracted a lot of attention due to their applications in 
spintronics \cite{rr1}, valleytronics \cite{rr2} and optoelectronics \cite{rr3, rrr10}. In this regard the group IV  
transition-metal dichalcogenides (TMDCs) have the form MX$_{2}$ (M=Mo,W; X=S,Se) are of particular interest due to their valley degree of freedom, large direct band gap \cite{rr3, rr4, rr5} and strong intrinsic spin-orbit interaction (SOI) \cite{rr6,rr7}.  Recently, nanoelectronic devices, such as amplifiers, photodetectors,  thin film transistors, and logical circuits \cite{rr8, rr9, rrr10, rr11}, based on their excellent electronic properties 
have been experimentally realized. In addition, several properties of TMDC monolayers have been investigated theoretically and experimentally \cite{rr12, rr13, rr14, rr15, rr16, rr17} e.g., 
magneto-optical spectra and magnetotransport.

Layered TMDCs, such as bilayer systems, 
exhibit a broad range of physical properties and have been extensively studied for applications in catalysis, tribology, electronics, photovoltaics, and electrochemistry \cite{rr18, rr19, rr20, rr21}. Also, few layer TMDCs have potential applications in nanoelectronics and nanophotonics. 
A field-effect transistor has been realized experimentally in a few-layer MoS$_{2}$ \cite{rr22}. Similarly, 
magnetoelectric effects and valley-controlled spin quantum gates \cite{rr23}, tuning of the valley magnetic moment \cite{rr24}, electrical control of the valley-Hall effect \cite{rr25}, and spin-layer locking effect \cite{rr26} has been explored in bilayer TMDCs. Most recently, magnetotransport studies of bilayer MoS$_{2}$ have been carried out \cite{rr27}. Additionally, a band gap tuning is  possible and more easily achievable in bilayer TMDCs than in monolayer TMDCs in the presence of a perpendicular  electric field $E_{z}$ \cite{rr28, rr29}. However, less attention has been paid to the optical properties of bilayer TMDCs in the simultaneous presence of  electric and magnetic fields.

In this work we study in detail the effect of magnetic and electric fields on the magneto-optical conductivity of  bilayer TMDCs with particular emphasis on the asymmetry between the conduction band (CB) and valence band (VB). Moreover, we assess the effect of the electric field on the band structure with and without magnetic field, and on the magneto-optical conductivities. Also, we compare our results with those for monolayer and bilayer graphene. 

We focus on bilayer WSe$_{2}$ due to recent experimental progress \cite{rr24, rr25, rr30, rr31, rr32, rr33} but our findings are equally pertinent 
to other bilayer TMDCs, e.g. MoSe$_{2}$ and WS$_{2}$. The 
WSe$_{2}$ bilayer has much stronger SOI in the conduction ($2\lambda_{c}=30$ meV) and valence ($2\lambda_{v}=450$ meV) bands compared  to bilayer MoS$_{2}$ ($2\lambda_{c}=0$ meV). The band-edge energy difference $E_{\Gamma K}$ between the $\Gamma$ and $K$ points is much smaller than in bilayer WSe$_{2}$ than in MoS$_{2}$ \cite{rr34, rr35}.  Therefore, the   CB and VB edges in bilayer WSe$_{2}$ lie at the $K$ point. Accordingly, bilayer WSe$_{2}$ has advantages over the MoS$_{2}$ when studying its optical properties due to the direct band gap at the $\pm K$ points.

In Sec. II we specify the Hamiltonian and obtain the energy eigenvalues and eigenfunctions with and without magnetic 
field. In Sec. III we present a general  expression for the conductivity $\sigma(\omega)$ and provide numerical results. Conclusions and a summary follow in Sec. IV.

\section{
 Energy spectrum}
In AB stacked bilayer TMDCs the top layer is 
rotated with respect to the bottom layer by 180 degrees 
such that the S atoms in it sit on top of the M atoms of the bottom layer. 
As a result, the effective Hamiltonian for bilayer TMDCs can be constructed from that of the single layer by simply adding the interlayer coupling term $\gamma$ \cite{rr36}. Then the one-electron Hamiltonian of bilayer WSe$_{2}$ near the $K$ and $K^{\prime}$ valleys reads \cite{rr23, rr24, rr26, rr37}
\begin{equation}
H^{\tau}=
\begin{pmatrix}
-\xi_{1}^{s \tau} && v_{F}\pi_{-}^{\tau} && \gamma && 0\\
v_{F}\pi_{+}^{\tau} && \xi_{2}^{s \tau} && 0 && 0\\
\gamma && 0 && -\xi_{3}^{s \tau} && v_{F}\pi_{+}^{\tau}\\
0 && 0 && v_{F}\pi_{-}^{\tau} && \xi_{4}^{s \tau}
\end{pmatrix}. \label{e1}
\end{equation}
Here $\tau=1 (-1)$ is for the $K$ ($K^{\prime}$) valley, $\pi_{\pm}^{\tau}=\tau\pi_{x}\pm i\pi_{y}$, $\xi_{1}^{s \tau}=\kappa+\tau s \lambda_{v} +s M_{z}-\tau M_{v}$, $\xi_{2}^{s \tau}=\alpha-\tau s \lambda_{c}-s M_{z}+\tau M_{v}$, $\xi_{3}^{s \tau}=\alpha-\tau s \lambda_{v}-s M_{z}+\tau M_{v}$, $\xi_{4}^{s \tau}=\kappa+\tau s \lambda_{c} + s M_{z}-\tau M_{v}$ and $\kappa=\Delta+V$ and $\alpha=\Delta-V$ with $\Delta$ the monolayer band gap. Further, $v_{F}= $5$\times$10$^{5}$ m/s is the Fermi velocity, $V$ is the potential difference between the two layers due to a perpendicular electric field $E_{z}$, and $\lambda$   the strength of the  SOI with spins up (down) represented by $s=+1(\uparrow)(s=-1(\downarrow))$.
Moreover, $M_{z}=g^{\prime}\mu_{B}B/2$ is the Zeeman exchange field induced by ferromagnetic order, $g^{\prime}$ the Land\'{e} $g$ factor $(g^{\prime}=g_{e}^{\prime}+g_{s}^{\prime})$, and $\mu_{B}$ the Bohr magneton \cite{rr38, rr39}; $g_{e}^{\prime}=2$ is the free electron $g$ factor and $g_{s}^{\prime}=0.21$ the out-of-plane factor due to the strong SOI. The term $M_{v}=g_{v}^{\prime} \mu_{B} B/2$ breaks the valley symmetry of the levels,  $g_{v}^{\prime}=4$ \cite{rr38, rr39}. The eigenvalues $E_{\mu}^{s,\tau}(k)$ of Eq. (1), when the magnetic field is absent, are
\begin{equation}
E_{\mu}^{s,\tau}(k)=\hslash v_{F} \varepsilon_{\mu}^{s,\tau}(k).\label{e2}
\end{equation}
The subscript $\mu=(\mu_{1},\mu_{2})$ is used for labeling  the energy bands: $\mu_{1}=+1 (-1)$ is for the electron (hole) branches and $\mu_{2}=+1(-1)$ is for the upper (lower) layer. Using the label $\mu_2$ is allowed provided the interlayer coupling is weak, see Refs. \onlinecite{rr23, rr26}. The factor $\varepsilon_{\mu}^{s,\tau}(k)\equiv\varepsilon$ in Eq. (\ref{e2}) is  the solution of the quartic equation 
\begin{eqnarray}
\notag
&\left[ \left(\varepsilon + \xi_{5}^{s \tau} \right) \left( \varepsilon - \xi_{6}^{s \tau}\right)-k^{2} \right] \left[ \left(\varepsilon + \xi_{7}^{s \tau}\right) \left( \varepsilon - \xi_{8}^{s \tau}\right)-k^{2} \right]  \\ &
-\gamma^{\prime 2} \left( \varepsilon - \xi_{6}^{s \tau}\right)\left( \varepsilon-\xi_{8}^{s \tau}\right)=0,\label{e3}
\end{eqnarray}
where $k\equiv k_y$ is the wave vector, $\varepsilon=E/\hslash v_{F}$, $\xi_{5}^{s \tau}=\xi_{1}^{s \tau}/\hslash v_{F}$, $\xi_{6}^{s \tau}=\xi_{2}^{s \tau}/ \hslash v_{F}$, $\xi_{7}^{s \tau}=\xi_{3}^{s \tau}/\hslash v_{F}$,  $\xi_{8}^{s \tau}=\xi_{4}^{s \tau}/ \hslash v_{F}$, and $\gamma^{\prime}=\gamma/\hslash v_{F}$. In the 
limit   $\xi_{i}^{s \tau}\rightarrow 0$, $i=5,..,8$, 
we obtain the  energy dispersion for bilayer graphene \cite{rr40}.

In the upper panel of Fig. 1 we plot the energy dispersion of bilayer WSe$_{2}$ for field $E_{z}=0$ at both valleys. We remark the following: (i) The splitting between the levels due to SOI is finite in the CB given by $2\lambda_{c}$ at $k=0$ in contrast to bilayer MoS$_{2}$ \cite{rr23, rr24, rr26, rr27}. Its means that four-fold degeneracy of CB in WSe$_2$ is partially lifted. So, the bands are two-two fold degenerate whereas it is four fold degenerate in bilayer MoS$_2$ at $k=0$. But, the splitting due to interlayer hopping is negligible in the CB.(ii) The value of interlayer hopping between the two layers is finite in the VB \cite{rr23, rr24, rr26, rr27}. So, splitting of levels in the VB is a combined effect of interlayer hopping and SOI given by $2[\lambda_{v}^{2}+\gamma^{2}]^{1/2}$ at $k=0$. This relation indicates that the VB is still split for $\lambda_{v}=0$ or $\gamma=0$. Further, levels in VB are also two-two fold degenerate as seen upper panel of Fig. (1). (iii) The gap between conduction and valence band edges is given by $2\Delta-\lambda_{c}-[\lambda_{v}^{2}+\gamma^{2}]^{1/2}$ for $k=0$.

For $E_{z}\neq 0$ we plot the energy spectrum in the lower panels of Fig. 1. We note the following: (i) The field $E_{z}$ modifies the SOI splitting. We note that two-fold spin degeneracy of all the bands in the CB and VB at each valley is completely lifted in contrast to bilayer MoS$_{2}$. However, bands have two-fold valley degeneracy i. e. energies of spin up and spin down bands at $K$ and $K^{\prime}$ valleys are same and vise versa.  (ii) An interlayer splitting is obtained in both the CB and VB. Analytically we obtain the gaps $2V\lambda_{v}/[\lambda_{v}^{2}+\gamma^{2}]^{1/2}$ for $V\ll \lambda_{v}$ and $2V$ at the valence and conduction bands edges, respectively. (iii) The band gap is also reduced by the field $E_{z}\propto V$. It is equal to $2\Delta-V-s\lambda_{c}-[\lambda_{v}^{2}+\gamma^{2}]^{1/2}-\tau s\lambda_{v} V/[\lambda_{v}^{2}+\gamma^{2}]^{1/2}$ for $V\ll \lambda_{v}$.
\begin{figure}[t]
\centering
\includegraphics[height=9cm, width=8cm]
{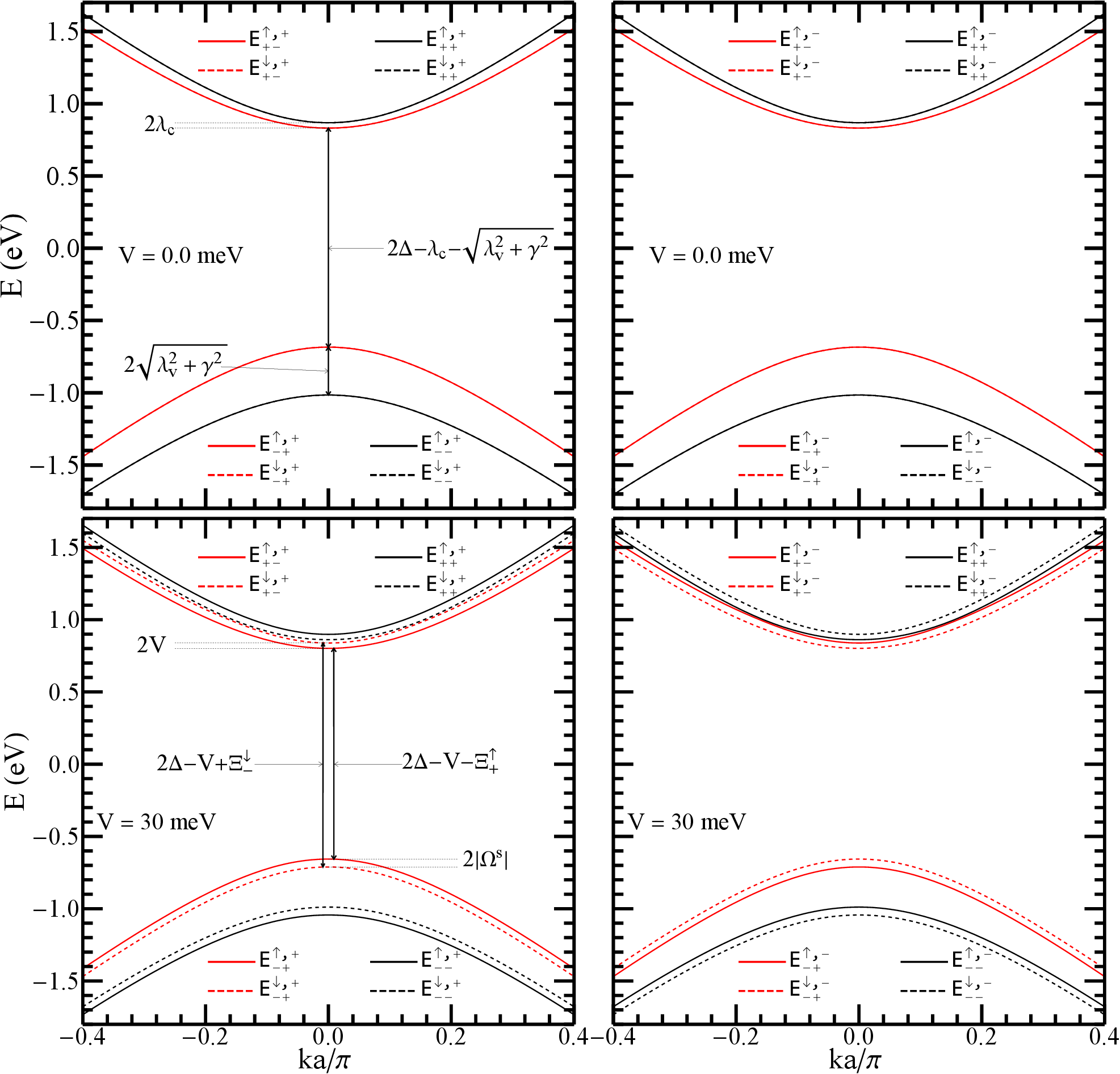}
\label{f4}


\caption{Band structure of bilayer WSe$_{2}$ for $2\lambda_{c}=37$ meV, $2\lambda_{v}=303$ meV, and $2\gamma=134$ meV. The upper panels are for 
$V=0$ meV,  the lower ones for $V=30$ meV. The left (right) panels are for the $K$ ($K^{\prime}$) valley and $\Xi_{\pm}^{s}= \lambda_{c} \pm \sqrt{\lambda_{v}^{2}+\gamma^{2}}+\Omega^{s}$ with $\Omega^{s}=s\lambda_{v} V/[\lambda_{v}^{2}+\gamma^{2}]^{1/2}$.}
\end{figure}
\subsection{Landau levels}

In the presence of a magnetic field $B$ perpendicular  to the layers we replace $\mathbf{\pi}$ by $-i\hslash\mathbf{\nabla}+\mathbf{A}$ in Eq. (\ref{e1}) and take the vector potential $\mathbf{A}$  in the Landau gauge $\mathbf{A}=(0,Bx,0)$. 
After diagonalizing Eq. (\ref{e1}) the Landau level (LL) spectrum is obtained as  
\begin{equation}
E_{n,\mu}^{s, \tau}=
\hslash \omega_{c}
\,\varepsilon_{n,\mu}^{s, \tau},\label{e4}
\end{equation}
with $\omega_{c}=v_{F}\sqrt{2eB/\hslash}$  the cyclotron frequency. 
For $n\geq1$ the factor $\varepsilon_{n,\mu}^{s, \tau}\equiv\varepsilon$ is the solution of the quartic equation
\begin{eqnarray}
\notag
&\left[ \left(\varepsilon+d_{1}^{s \tau}\right) \left( \varepsilon - d_{2}^{s \tau}\right)-n \right] \left[ \left(\varepsilon+d_{3}^{s \tau}\right) \left( \varepsilon- d_{4}^{s \tau}\right)-(n+1) \right]
\\&
 - t^{2} \left( \varepsilon-d_{2}^{s \tau}\right)\left( \varepsilon- d_{4}^{s \tau}\right)=0,\label{e5}
\end{eqnarray}
where $t = \gamma /\hslash \omega_{c}$, $d_{1}^{s \tau}=\kappa^{\tau}+s\lambda_{v}+\tau(sM_{z}-\tau M_{v})/\hslash \omega_{c}$, $d_{2}^{s \tau}=\alpha^{\tau}-s\lambda_{c}-\tau(sM_{z}-\tau M_{v}) /\hslash \omega_{c}$, $d_{3}^{s \tau}=\alpha^{\tau}-s\lambda_{v}-\tau(sM_{z}-\tau M_{v}) /\hslash \omega_{c}$, and $d_{4}^{s \tau}=\kappa^{\tau}+s\lambda_{c}+\tau(sM_{z}-\tau M_{v}) /\hslash \omega_{c}$ with $\kappa^{\tau}=\Delta+\tau V$ and $\alpha^{\tau}=\Delta-\tau V$. 
In the 
limit   $\xi_{i}^{s \tau}\rightarrow 0$, $i=5,..,8$, 
Eq. (\ref{e4}) gives a LL dispersion similar to  that of bilayer graphene  \cite{ rr41, rr42}. The eigenfunctions are
\begin{eqnarray}
\notag
&\psi_{n,\mu}^{s,+}&=
\frac{1}{\sqrt{L_{y}}}
\begin{pmatrix}
\varrho_{n,\mu}^{s,+} \phi_{n} \\
\vspace*{0.2cm}
\Theta_{n,\mu}^{s,+}\, \phi_{n-1}\\
\vspace{0.2cm}
\Lambda_{n,\mu}^{s,+}\,\phi_{n}\\
\vspace{0.2cm}%
\Upsilon_{n,\mu}^{s,+}\,\phi_{n+1}
\end{pmatrix}
e^{ik_{y}y}\,, \quad \quad 
\vspace{2cm}
\\&
\psi_{n,\mu}^{s,-}&=
\frac{1}{\sqrt{L_{y}}}
\begin{pmatrix}
\Lambda_{n,\mu}^{s,-}\,\phi_{n}\\
\vspace{0.2cm}
\Upsilon_{n,\mu}^{s,-}\,\phi_{n+1}\\
\vspace{0.2cm}
\varrho_{n,\mu}^{s,-}\, \phi_{n}\\
\vspace{0.2cm}
\Theta_{n,\mu}^{s,-}\, \phi_{n-1}
\end{pmatrix}
e^{ik_{y}y}.
\label{e6}
\end{eqnarray}
Here $\phi_{n}\equiv \phi_{n}(\upsilon)= (2^{n}n!l_{B} \sqrt{\pi})^{-1/2} e^{-\upsilon^{2}/2} H_{n}(\upsilon)$ is the harmonic oscillator wave function with $\upsilon=(x-l_{B}^{2}k_{y})/l_{B}$ and $H_{n}(\upsilon)$ the Hermite polynomial of order $n$.  Notice that $\phi_{n}\equiv 0$ for $n<0$. The coefficients are given by $\Theta_{n,\mu}^{s,\tau}=
\sqrt{n}\,\varrho_{n,\mu}^{s,\tau}/[\varepsilon_{n,\mu}^{s,\tau}-d_{2}^{s \tau}] $, $\Lambda_{n,\mu}^{s,\tau}=k_{n,\mu}^{s,\tau}\varrho_{n,\mu}^{s,\tau}$, and $\Upsilon_{n,\mu}^{s,\tau}=
\sqrt{n+1}\,k_{n,\mu}^{s,\tau}\,\varrho_{n,\mu}^{s,\tau}/[\varepsilon_{n,\mu}^{s,\tau}-d_{4}^{s \tau}]$,
with $\varrho_{n,\mu}^{s,\tau}$ the normalization constants
\begin{eqnarray}
\hspace*{-0.2cm}
&\varrho_{n,\mu}^{s, \tau}= \Big\{(k_{n,\mu}^{s, \tau})^{2}\big[1+
\frac{(n+1)}{\big(\varepsilon_{n,\mu}^{s, \tau}-d_{4}^{s  \tau}\big)^{2}}\big]+1
+\frac{n}{\big(\varepsilon_{n,\mu}^{s, \tau}-d_{2}^{s \tau}\big)^{2}} \Big\}^{-1/2}\label{e7}
\end{eqnarray} 
and $k_{n,\mu}^{s, \tau}=[(\varepsilon_{n,\mu}^{s, \tau}+d_{1}^{s \tau})(\varepsilon_{n,\mu}^{s, \tau}-d_{2}^{s  \tau})-n]/t(\varepsilon_{n,\mu}^{s,\tau}-d_{2}^{s  \tau})$. As Eq. (\ref{e6}) shows, the full wave function 
is a mixture of the Landau wave functions with indices $n-1$, $n$, and $n+1$. 

For $n=0$ there are two special LLs. 
One has the ener-\\gies $\varepsilon_{0,+-}^{s,+}=d_{4}^{s+}$ and $\varepsilon_{0,+-}^{s,-}=d_{2}^{s-}$ for the $K$ and $K^{\prime}$ val-\\leys, respectively. 
The corresponding 
wave functions are 
\begin{equation}
\psi_{0,+-}^{s,+}=
\frac{1}{\sqrt{L_{y}}}
\begin{pmatrix}
0\\
0\\
0\\
\phi_{0}
\end{pmatrix}
e^{ik_{y}y}\,, \quad \quad %
\label{e8}
\hspace{-0.5cm}
\psi_{0,+-}^{s,-}=
\frac{1}{\sqrt{L_{y}}}
\begin{pmatrix}
0\\
\phi_{0}\\
0\\
0
\end{pmatrix}
e^{ik_{y}y}.
\end{equation}
This LL 
has exactly the same properties as  the $n=0$ conventional, non-relativistic LL. For $\Delta=\lambda_{c}=\lambda_{v}=V=0$, this level has exactly  zero energy as the $n=0$ LL for bilayer graphene \cite{ rr41, rr42}. Also, from  Eq. (\ref{e5}) we obtain three other levels for $n=0$. We obtain the wave functions for two of these levels from Eq. (\ref{e6}) by simply setting $n=0$ in it. Further, we specify the quantum number $(\mu)$ labels for these two levels as  $\mu=(+,+)$ and $\mu=(-,+)$. However, for the third LL we specify 
$n$ and $\mu$ as  shown in the eigenfunctions
\begin{eqnarray}
\notag
&
\psi_{0,--}^{s,+}&=
\frac{1}{\sqrt{L_{y}}}
\begin{pmatrix}
\varrho_{0,--}^{s,+}\phi_{0}\\
0\\
\Lambda _{0,--}^{s,+}\phi_{0}\\
\varrho_{0,--}^{s,+} t \phi_{1}
\end{pmatrix}
e^{ik_{y}y}, 
%
\\&
\psi_{0,--}^{s,-}&=
\frac{1}{\sqrt{L_{y}}}
\begin{pmatrix}
 \Lambda_{0,--}^{s,-}\phi_{0} \\
 \varrho_{0,--}^{s,-} t\phi_{1}\\
\varrho_{0,--}^{s,-}\phi_{0}\\
0
\end{pmatrix}
e^{ik_{y}y},
\label{e9}
\end{eqnarray}
where $\Lambda_{0,--}^{s,\tau}=\varrho_{0,--}^{s,\tau} t (\varepsilon_{0,--}^{s,\tau}-d_{4}^{s,\tau})$. The normalization constants are 
\begin{equation}
\varrho_{0,--}^{s,\tau}=k_{0,--}^{s,\tau} \big \{(k_{0,--}^{s,\tau})^{2} + t^{2}[1+(\varepsilon_{0,--}^{s,\tau}-d_{4,--}^{s,\tau})] \big \}^{-1/2}\label{e10}
\end{equation}
and $k_{0,--}^{s,\tau}=(\varepsilon_{0,--}^{s,\tau}+d_{3,--}^{s,\tau}) (\varepsilon_{0,--}^{s,\tau} - d_{4,--}^{s,\tau})-1$. The wave function corresponding to this LL is a mixture of the $n = 0$ and $n = 1$ conventional (nonrelativistic) Landau functions $\phi_{0}$ and $\phi_{1}$. For $\Delta=\lambda_{c}=\lambda_{v}=V=0$  Eq. (\ref{e9}) gives  
the eigenfunctions
   for bilayer graphene \cite{ rr41, rr42}.
\begin{figure}[t]
\centering

\hspace*{0.7cm}
\includegraphics[height=1.5cm, width=8cm] 
{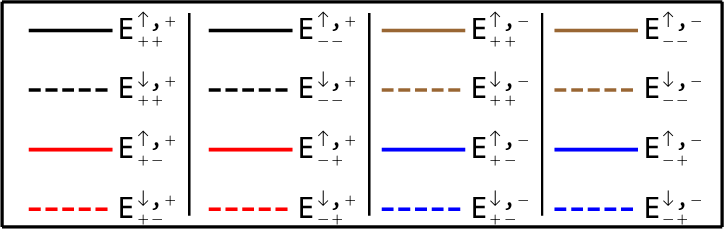}
\ \\

\vspace*{0.3cm}

\includegraphics[height=12cm, width=9cm]
{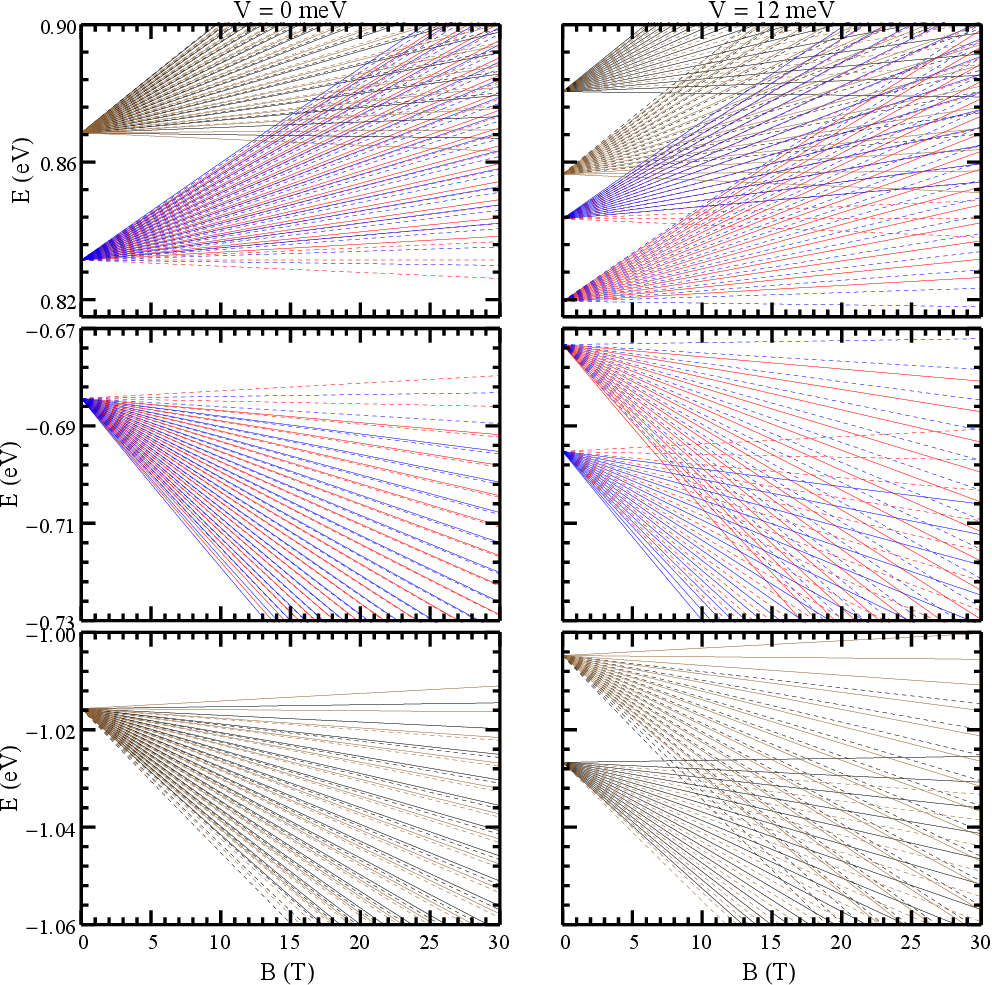}


\caption{Energy spectrum of bilayer WSe$_{2}$ versus magnetic field $B$ for $M_{z}\neq, M_{v}\neq 0$. The left panel is for the $E_{z}=0$ and right one for $E_{z}\neq 0$, respectively. The upper panel explains the colour and style assignments of the curves.}
\label{f2}
\end{figure}

In Fig. 2 (left panels) we plot the spin and valley dependent LL spectrum, but independent of $k_{y}$, given by Eq. (4) versus the magnetic field $B$ for $V=0$ and finite spin $M_{z}$ and valley $M_{v}$ Zeeman fields. The marking of all  curves is explained in the upper panel.
We find the following: (i) The energy spectrum grows linearly with $B$ due to the huge band gap. (ii) For $M_{z}=M_{v}=0$,  all LLs ($n\geq 1$) are two-fold degenerate corresponding to the two valleys including the $n=0$ LL with energies $\varepsilon_{0,++}^{s,\tau}$, $\varepsilon_{0,-+}^{\downarrow,+}$, $\varepsilon_{0,-+}^{\uparrow,-}$, and $\varepsilon_{0,--}^{\downarrow,\tau}$ in both the conduction and valence bands. The LL with energy $\varepsilon_{0,+-}^{s,\tau}=\Delta+ s\lambda_{c}$ for $n=0$ is doubly degenerate in the conduction band, i.e. $\varepsilon_{0,+-}^{\uparrow,+}\equiv \varepsilon_{0,+-}^{\uparrow,-}$ and $\varepsilon_{0,+-}^{\downarrow,+}\equiv \varepsilon_{0,+-}^{\downarrow,-}$. Further, in the valence band the $n=0$ LL is two-fold valley degenerate, i.e. $\varepsilon_{0,-+}^{\uparrow,+}\equiv \varepsilon_{0,-+}^{\downarrow,-}$ and  $\varepsilon_{0,--}^{\uparrow,+}\equiv \varepsilon_{0,--}^{\uparrow,-}$. In this situation, interlayer splitting among the levels of WSe$_{2}$ or MoS$_{2}$ bilayer is zero \cite{rr27}. On the other hand, the intra-layer spin splitting in bilayer WSe$_{2}$ is significantly large given by $2\lambda_{c}$, which can be clearly seen in the limit of vanishing $B$ as compared to bilayer MoS$_{2}$ \cite{rr27}. (iii) For $M_{z}\neq 0, M_{v}\neq 0$, shown in the left panel of Fig. 2, the spin and valley degeneracies of all LLs $(n\geq 0)$ are lifted i.e., the energies of the spin-up (-down) LLs at the $K$ valley are different than  the spin-down (-up) ones at the $K^{\prime}$ valley in contrast to the $B=0$ case. (iv) The valley Zeeman term $M_{v}$ lifts the spin degeneracy as well as the valley degeneracy in both the conduction and valence bands. This 
effect on the LLs, due to the $M_{v}$ term,  is absent in bilayer MoS$_{2}$ \cite{rr27}. Notice that the inter-layer splitting among the levels of  bilayer WSe$_{2}$ vanishes 
in contrast to bilayer MoS$_{2}$ \cite{rr27}.

We show the LL spectrum in Fig. 2 (right panels) for finite field $E_{z}$ ($V=12$ meV) including the $M_{z}$ and $M_{v}$ terms. We deduce the following: (i) The field $E_{z}$ modifies the interlayer splitting, e.g., it makes it $24$ meV and $23$ meV in the conduction and valence bands, respectively. (ii) The spin and valley degeneracies of all levels ($n\geq 0$) are completely lifted, i.e., the energies of the spin-up $(\uparrow)$ states at the $K$ valley and a spin-down $(\downarrow)$ ones at the $K^{\prime}$ valley are totally different in contrast to the $B=0$ case. Moreover, we can adjust the LL separation  by varying the external electric and magnetic fields. This  becomes important when we tune the onset frequency of the magneto-optical conductivity.
\begin{figure}[t]
\centering

\includegraphics[height=8cm, width=8cm] 
{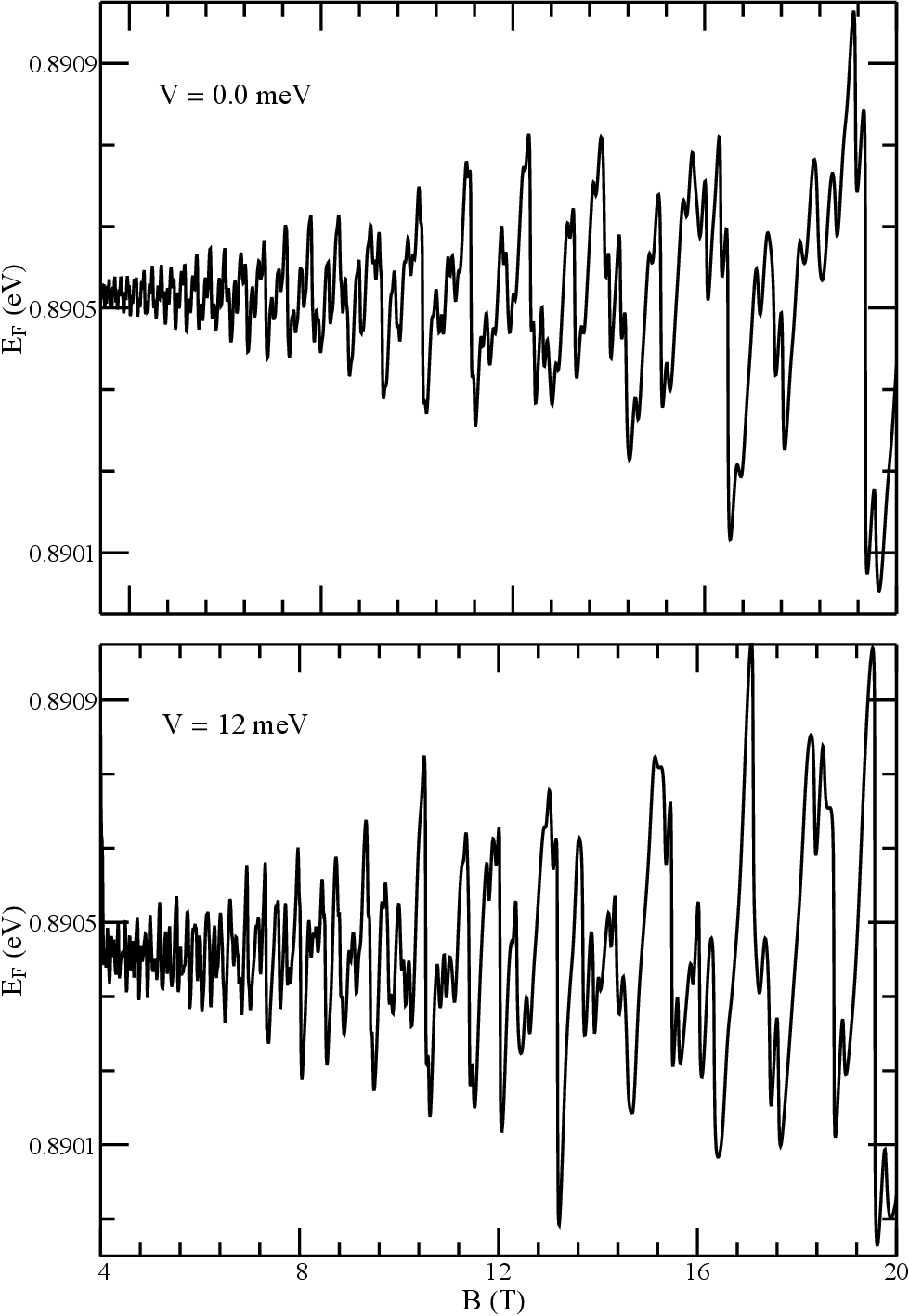}
\vspace*{-0.3cm}
\caption{Fermi energy $E_{F}$ versus $B$ for an electron density $n_{e}=4.3 \times 10^{13}$ cm$^{-2}$. The upper panels are for $V=0$ meV and the lower ones for $V=12$ meV. }

\label{ff5}
\end{figure}
\subsection{Density of states}

The density of states $D(E)$ is given by 
\begin{equation}
D(E)=\dfrac{1}{S_{0}}\sum_{n,\tau,s,\mu,k_{y}} \delta(E-E_{n,\mu}^{s,  \tau}),\label{e11}
\end{equation}
where $S_{0}=L_{x}L_{y}$ is the area of the system. The sum over $k_{y}$ can be calculated by using $k_{0}=L_{x}/2l_{B}^{2}$ and the prescription $\sum_{k_{y}}\rightarrow (L_{y}/2\pi)g_{s}g_{v} \int_{-k_{0}}^{k_{0}} dk_{y}=(S_{0}/D_{0})g_{s}g_{v}$, with $D_{0}=2\pi l_{B}^{2}$; $g_{s} (g_v)$ denotes the spin (valley) degeneracy. In this work we take $g_{s}=g_{v}=1$ because the spin and valley degeneracies are lifted.  $E_F$ at constant electron concentration $n_{e}$ we  obtain $E_F$  from the relation
\begin{equation}
n_{e}= \int_{-\infty}^{\infty} D(E)f(E) dE=\dfrac{g_{s\slash v}}{D_{0}}\sum_{n,\tau,s,\mu}f(E_{n,\mu}^{s,  \tau}),\label{e12}
\end{equation}
where $f(E_{n,\mu}^{s, \tau})=1/\big[1+\exp[\beta(E_{n,\mu}^{s, \tau}-E_{F})]\big]$ is the Fermi-Dirac function and $\beta=1/k_{B}T$. 

The black solid curve in the upper panels of Fig. 3 shows $E_{F}$, obtained from Eq. (\ref{e12}) numerically, versus $B$ for $E_{z}=0$. The  field $B$ lifts the spin and valley degeneracies 
 of all  LLs $(n\geq 0)$, i.e. the spin-up and  spin-down electrons in the $K$ valley have  different energies than the corresponding 
 ones in the $K^{\prime}$ valley.
 This leads to additional intra-LL small jumps in Fig. 3 (upper panels) that are 
  enhanced, as shown in the lower panels of Fig. 3, when a finite electric field $E_{z}$ is applied.
\begin{figure}[t]
\centering

\includegraphics[height=8cm, width=8cm]
{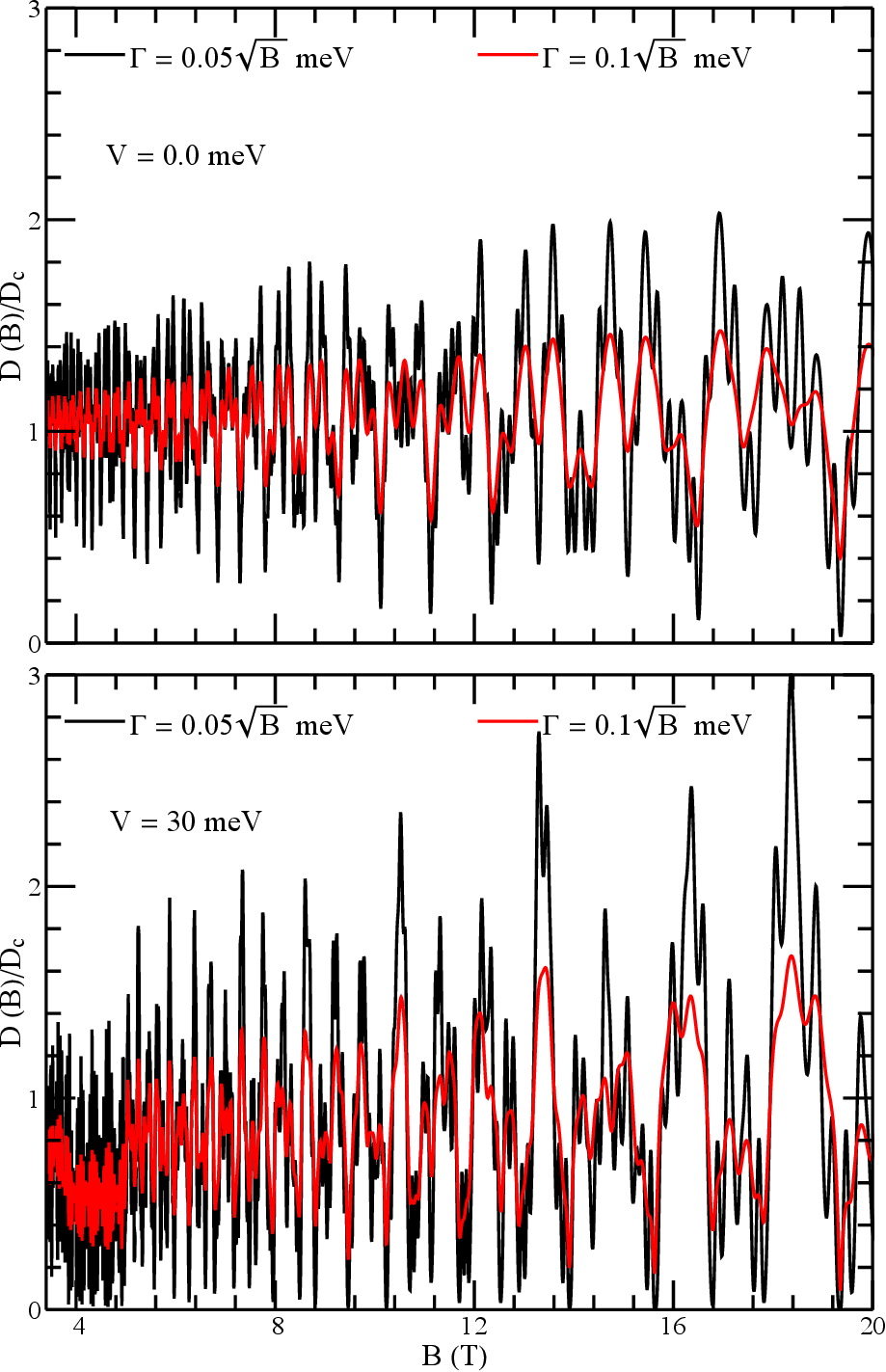}

\vspace*{-0.3cm}

\caption{Dimensionless density of states as a function of field $B$ for LL width $\Gamma=0.05\sqrt{B}$ meV (black curve) and $\Gamma=0.1\sqrt{B}$ meV (red curve). The upper panels are for $V=0$ and the lower ones for $V=30$ meV. The left and right panels differ only in the magnetic field range ($x $ axis). }
\label{f5}
\end{figure}

We evaluate $D(E)$ per unit area  assuming a Gaussian broadening of the $\delta$ function in Eq. (\ref{e11}). At zero tempe-\\rature we have 
$D(E)=(g_{s}g_{v}/D_{0}\Gamma\sqrt{2\pi})\sum_{\zeta}\exp[-(E-E_{\zeta})^{2}/2\Gamma^{2}]$, where $\Gamma$ is the width of the 
distribution and $\vert \zeta \rangle \equiv \vert n,\mu,s,\tau,k_{y}\rangle$. In Fig. 4 we plot the dimensionless $D(E)$ versus the field $B$ in the conduction band for two different values of $E_{z}$ and $\Gamma$. The Shubnikov-de Haas (SdH) oscillations are clearly shown. 
The level broadening effect becomes significant for weak $B$ fields due to the small LL separation. On the other hand, this effect may become very weak in strong 
fields $B$ for which the  LL separation is strong and  $\Gamma\propto \sqrt{B}$.  

Looking closely at Fig. 4 we observe a beating of the SdH oscillations at low fields $B$ and a pronounced splitting at higher fields. 
The beating of the oscillations is observed  
  for $B \leq 10$ T, with $E_{z} = 0$, and for $ B \leq 5$ T with $E_{z}\neq 0$. Away from these ranges the beating pattern is replaced by a split in the SdH oscillations. This behaviour is explained by the closeness of the oscillation frequencies of the SOI-split LLs. The  field $B$ enhances the splitting in the conduction band by mixing the spin-up and spin-down states of neighbouring LLs into two unequally spaced energy branches. This is also the case of a 2DEG
\cite{rr43}. This 
beating  pattern occurs when the level broadening is of the order of 
$\hslash \omega_{c}$; it is replaced by a split in the oscillations when the SOI becomes weak for large fields B. We further notice 
that the beating pattern shifts to lower magnetic fields  for finite 
electric field energy $V$.
\begin{figure}[t]
\centering

\vspace*{-0.2cm}
\includegraphics[width=.45\textwidth]{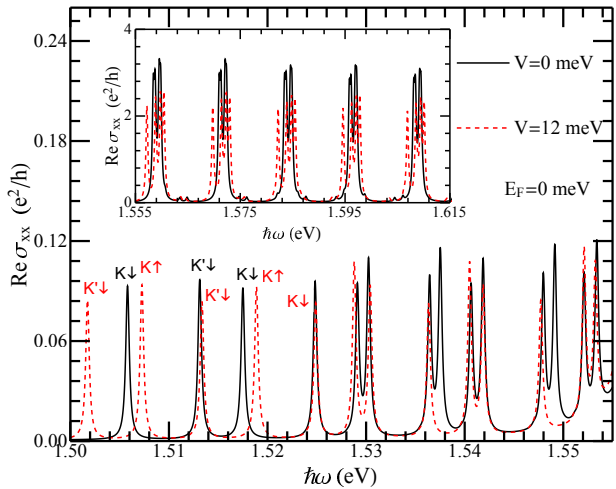}

\vspace*{-0.3cm}

\caption{Real part of the longitudinal optical conductivity $\sigma_{xx}^{nd}(\omega)$ versus 
the photon energy $\hslash \omega$ for a field $B=30$ T. The solid black and dotted red curves are for $V=0$ and $V=12$ meV, respectively. The inset shows Re$\sigma_{xx}^{nd}(\omega)$ 
for higher $\hslash \omega$. The spin assignment of the curves follows from Eq. (\ref{16}).}
\label{f6}
\end{figure}

\section{
Conductivities}

We consider a many-body system described by the Hamiltonian $H = H_{0} + H_{I} - \mathbf{R \cdot F}(t)$, where $H_{0}$ is the unperturbed part, $H_{I}$ is a binary-type interaction (e.g., between electrons and impurities or phonons), and $ \mathbf{- R \cdot F}(t)$ is the interaction of the system with the external field F(t) \cite{rr44}. For conductivity problems we have $\mathbf{F}(t) = e \mathbf{E}(t)$, where $\mathbf{E}(t)$ is the electric field, $e$ the electron charge, $\mathbf{R = \sum_{r_{i}}}$ , and $\mathbf{r_{i}}$  the position operator of electron $i$. In the representation in which $H_{0}$ is diagonal the many-body density operator $\rho = \rho^{d} + \rho^{nd}$ has a diagonal part $\rho^{d}$ and a nondiagonal part $\rho^{nd}$. For weak electric fields and weak scattering potentials, for which the first Born approximation applies, the conductivity tensor has a diagonal part $\sigma_{\mu\nu}^{d}$ and a nondiagonal part $\sigma_{\mu\nu}^{nd}$ , $\sigma_{\mu\nu} = \sigma_{\mu\nu}^{d} + \sigma_{\mu\nu}^{nd}, \mu,\nu = x,y$. 

In general we have two kinds of currents, diffusive and hopping, with $\sigma_{\mu\nu}^{d} = \sigma_{\mu\nu}^{dif} + \sigma_{\mu\nu}^{col}$, but usually only one of them is present. When a magnetic field is present we have only a hopping current since the diffusive part $\sigma_{\mu\nu}^{dif}$ vanishes identically due to the vanishing velocity matrix elements as is evident, for 
elastic scattering, by its form \cite{rr44}
\begin{equation}
\sigma_{\mu\nu}^{d} (\omega) = \dfrac{\beta e^{2}}{S_{0}} \sum_{\zeta} f_{\zeta} (1 - f_{\zeta} ) \dfrac{v_{\nu\zeta} v_{\mu\zeta} \tau_{\zeta}}{1 + i\omega \tau_{\zeta}} , \label{l1}
\end{equation}
where $\tau_{\zeta}$ is the momentum relaxation time, $\omega$ the frequency, and $v_{\mu\zeta}$ the diagonal matrix elements of the velocity operator. Further, $f_{\zeta} = [1 + \exp \beta (E_{\zeta} - E_{F})]^{-1}$ is the Fermi-Dirac distribution function, $\beta = 1/k_{B}T$, $T$ the temperature,  and $S_{0}$ the area of the sample. In our case $v_{\mu\zeta} = 0$ and the conductivity given by Eq. (\ref{l1}) vanishes. 

The ac hopping conductivity $\sigma_{\mu\nu}^{col}(\omega)$  is given by Eq. (2.64) of Ref. [44]. In strong fields $B$ it is much smaller than the contribution $\sigma_{\mu\nu}^{nd}$, given below, and is neglected. 
\begin{figure}[t]
\centering
\includegraphics[height=5cm, width=8cm]
{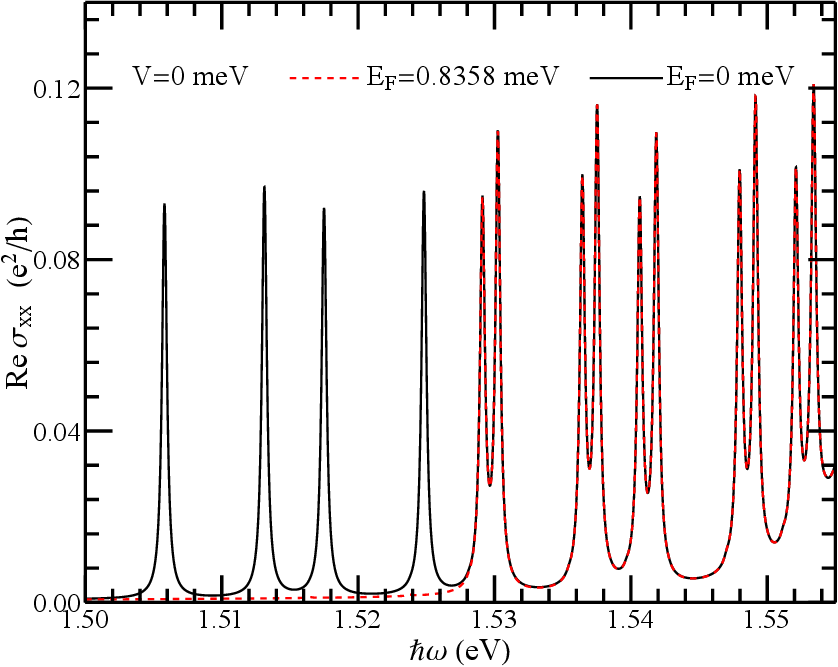}

\vspace*{-0.3cm}

\caption{As in Fig. 5 but for 
two different values of $E_{F}$ as indicated.} 
\label{f7}
\end{figure}
Regarding this contribution $\sigma_{\mu\nu}^{nd}$ one can use the identity $f_{\zeta} (1 - f_{\zeta^{\prime}})[1 - \exp \beta (E_{\zeta} - E_{\zeta^{\prime}})] = f_{\zeta} - f_{\zeta^{\prime}}$ and cast the original form in the more familiar one \cite{rr44}

\begin{equation}
\sigma_{\mu\nu}^{nd} (\omega) =\dfrac{ i\hslash e^{2}}{S_{0}}\sum_{\zeta \neq \zeta^{\prime}} \dfrac{(f_{\zeta} - f_{\zeta^{\prime}}) v_{\nu\zeta\zeta^{\prime}} v_{\mu\zeta\zeta^{\prime}}}{(E_{\zeta} - E_{\zeta^{\prime}})(E_{\zeta} - E_{\zeta^{\prime}} + \hslash \omega - i \Gamma )} ,\label{l2}
\end{equation}
where the sum runs over all quantum numbers $\vert \zeta \rangle \equiv \vert n,\mu,s,\tau,k_{y}\rangle$ and $\vert \zeta^{\prime} \rangle \equiv \vert n^{\prime},\mu^{\prime},s^{\prime},\tau^{\prime},k_{y}^{\prime}\rangle$ with $\zeta \neq \zeta^{\prime}$. The infinitesimal quantity $\epsilon$ in the original form \cite{rr44} has been replaced by $\Gamma_{\zeta}\approx \Gamma$ to account for the broadening of the energy levels. The familiar selection rules $n^{\prime}=n\pm 1$ are obtained through an evaluation of velocity matrix elements, see  Eqs. (\ref{l4})- (\ref{l5}) below. In the zero-temperature limit the Fermi function can be replaced by a step function.  Further, we assume positive values of  $E_{F}$, so that all  transitions to negative levels are Pauli blocked. In Eq. (14) $v_{\nu \zeta \zeta^{\prime}}$ and $v_{\mu \zeta \zeta^{\prime}}$ are the off-diagonal matrix elements of the velocity operator. They are evaluated using the operator expressions $v_{x}=\partial H/\partial p_{x}$ and $v_{y}=\partial H/\partial p_{y},$ and are given in terms of the Pauli matrices $\sigma_{\upsilon}$ as 
\begin{align}
\begin{split}
v_{x}=
\tau v_{F}
\begin{pmatrix}
\sigma_{x} && 0\\
0 && \sigma_{x}
\end{pmatrix}
,\ \ \ \ v_{y}=
v_{F}
\begin{pmatrix}
\sigma_{y} && 0\\
0 && -\sigma_{y}
\end{pmatrix},\label{l3}
\end{split}
\end{align}
%
\begin{figure}[t]
\centering

\includegraphics[height=4.5cm, width=8cm]
{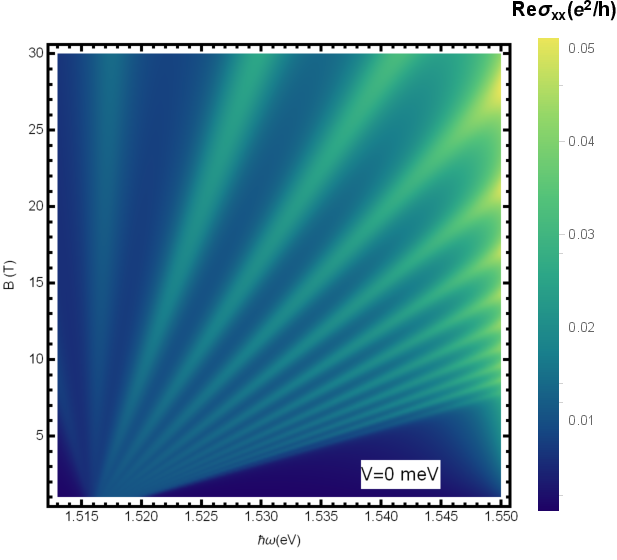}
\includegraphics[height=4.5cm, width=8cm]
{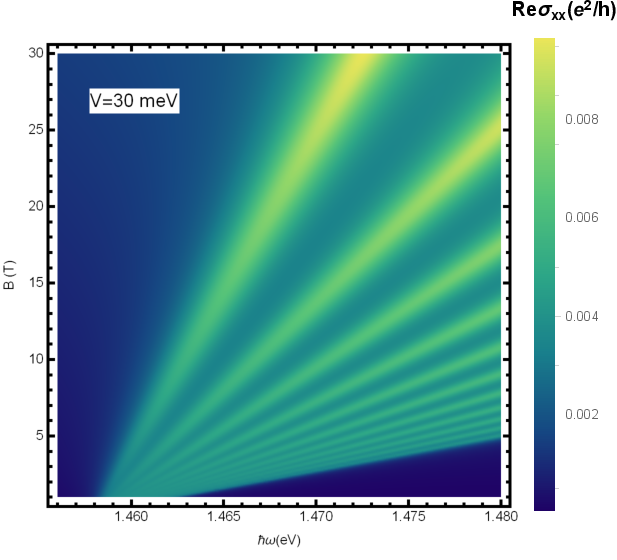}


\caption{ ($B, \omega$) Contour 
plot 
of the real part of the longitudinal conductivity for 
 $E_{z}=0$ (upper panel) and $E_{z} \neq 0$ (lower panel). The level width $\Gamma$ is set to $0.4 \sqrt{B}$ meV.}
\label{f16}
\end{figure}

With $\varepsilon_{n,d_{2}}\equiv\varepsilon_{n,\mu}^{s ,\tau}- d_{2}^{s \tau}$  , $\varepsilon_{n,d_{4}}\equiv\varepsilon_{n,\mu}^{s , \tau}- d_{4}^{s \tau}$ and $Q=v_{F}\varrho_{n,\mu}^{s , \tau} \varrho_{n^{\prime},\mu^{\prime}}^{s^{\prime} , \tau^{\prime}}\,\delta_{s,s^{\prime}}$, and $R=k_{n,\mu}^{s, \tau}\ k_{n^{\prime},\mu^{\prime}}^{s^{\prime},\tau^{\prime}}$ the results are
\begin{widetext}
\begin{eqnarray}
\left\langle \zeta 
\right\vert v_{x}\left\vert \zeta^\prime 
\right\rangle   =\tau Q 
  \Big[ \sqrt{n+1}\ 
  \Big(  
   \frac{ 1
}{  \ \varepsilon_{n, d_{2}}^\prime   }  
+\frac{
R
}{\varepsilon_{n,d_{4}}  }     \Big)\delta_{n,n^{\prime}-1} 
+  \sqrt{n} \ \Big(    \frac{ 1
}{  \ \varepsilon_{n,d_{2}}   }  +\frac{
R
}{\varepsilon_{n,d_{4}}^{\prime}  }     \Big)\delta_{n,n^{\prime}+1} \Big],\label{l4}
\end{eqnarray}
\begin{eqnarray}
\left\langle  \zeta^\prime 
\right\vert
v_{y}\left\vert  \zeta
\right\rangle   
=\tau iQ
  \Big[ \sqrt{n+1}\ \Big(    \frac{ 1
}{  \ \varepsilon_{n,d_{2}}^\prime   }  +\frac{
R
}{\varepsilon_{n,d_{4}}  }     \Big)\delta_{n,n^{\prime}-1} 
- \sqrt{n} \ \Big(    \frac{ 1
}{  \ \varepsilon_{n,d_{2}}   }  +\frac{
R
}{\varepsilon_{n,d_{4}}^{\prime}  }     \Big)\delta_{n,n^{\prime}+1} \Big],\label{l5}
\end{eqnarray}
where $\mu=\left\{  \mu_{1},\mu_{2}\right\}$. Using Eqs. (\ref{l4}), (\ref{l5}), and (\ref{l2}) we obtain the real and imaginary 
parts of the conductivities $\sigma_{xx}^{nd}(\omega)$ and $\sigma_{xy}^{nd}(\omega)$ which for convenience and later purposes we write, setting                     $\Delta_{n,n+1}= \varepsilon_{n,\mu}^{s, \tau}-\varepsilon_{n+1,\mu}^{s, \tau}$, as

\begin{align}
\begin{split}
\begin{pmatrix}
\mathrm{Re}\sigma_{xx}^{nd}\\
\ \\
\mathrm{Im} \sigma_{xy}^{nd}
\end{pmatrix} =\mp\frac{e^{2}}{2h} \sum_{s,\tau,n,\mu,\mu^{\prime}}
\eta_{n,\mu,\mu^{\prime}}^{s, \tau}\bar{\Gamma} \Big[ \dfrac{1}{\bigl(\Delta_{n,n+1}+\bar{\omega} \bigr)^{2}+\bar{\Gamma} ^{2}} \pm \dfrac{1}{\bigl(\Delta_{n,n+1}-\bar{\omega} \bigr)^{2}+\bar{\Gamma}^{2}}  \Big], \label{16}
\end{split}
\end{align}
\begin{align}
\begin{split}
\begin{pmatrix}
\mathrm{Im}\sigma_{xx}^{nd}\\
\ \\
\mathrm{Re}\sigma_{xy}^{nd}
\end{pmatrix} =-\frac{e^{2}}{2h} \sum_{s,\tau,n,\mu,\mu^{\prime}}
\eta_{n,\mu,\mu^{\prime}}^{s, \tau}\  \Big[ \dfrac{\Delta_{n,n+1}+\bar{\omega}}{\bigl(\Delta_{n,n+1}+\bar{\omega} \bigr)^{2}+\bar{\Gamma} ^{2}} \mp \dfrac{\Delta_{n,n+1}-\bar{\omega}}{\bigl(\Delta_{n,n+1}-\bar{\omega} \bigr)^{2}+\bar{\Gamma}^{2}}  \Big], \label{166}
\end{split}
\end{align}
with
\begin{align}
\begin{split}
&\eta_{n,\mu,\mu^{\prime}}^{s, \tau}=(
n+1) 
 \big(  \varrho_{n,\mu}^{s, \tau}\varrho_{n+1,\mu^{\prime}}^{s, \tau}\big)^{2} \,  \Big[  \frac{
    k_{n,\mu}^{s, \tau}\ k_{n+1,\mu^{\prime}}^{s, \tau}}      
  {  \ \varepsilon_{n,d_{4}}     }          +         \frac{1}{\varepsilon_{n+1,d_{2}}  } \Big]  ^{2}
    \frac{
\ \ f(E_{n,\mu}^{s,\tau})-f(E_{n+1,\mu^{\prime}}^{s,\tau}) 
}{  \varepsilon_{n,\mu}^{s, \tau}- \varepsilon_{n+1,\mu^{\prime}}^{s, \tau
}  }. \label{l7}
\end{split}
\end{align}
\end{widetext}
Here $\bar{\omega}\equiv\omega/\omega_{c}$ and $\bar{\Gamma}\equiv \Gamma/\hslash\omega_{c}$. The Fermi Dirac function at $T$ = 0 becomes  
the Heaviside step function $\Theta(x)$ and enforces the Pauli exclusion principle for optical transitions, i.e., transitions  occur only between the occupied $n$  state and the unoccupied $n^{\prime}$ one. The $n =0$ contributions to the absorptive conductivity  Eq. (\ref{16}) are evaluated separately. The results are given by Eq. (\ref{A1}) in  Appendix A.

Notice that in the limit $\omega\to 0, \Gamma\to 0$ we have 
\begin{align}
\mathrm{Re}\ \sigma_{xx}^{nd}=\mathrm{Im}\ \sigma_{xx}^{nd}=
\mathrm{Im}\ \sigma_{xy}^{nd}
 = 0, \label{19}
\end{align}
\begin{align}
\mathrm{Re}\ \sigma_{xy}^{nd}
 = - \frac{e^{2}}{h} \sum_{s,\tau,n,\mu,\mu^{\prime}}
  \dfrac{\eta_{n,\mu,\mu^{\prime}}^{s, \tau} }{\Delta_{n,n+1}}. \label{110}
\end{align}
\begin{figure}[t]
\centering

\includegraphics[width=.47\textwidth]{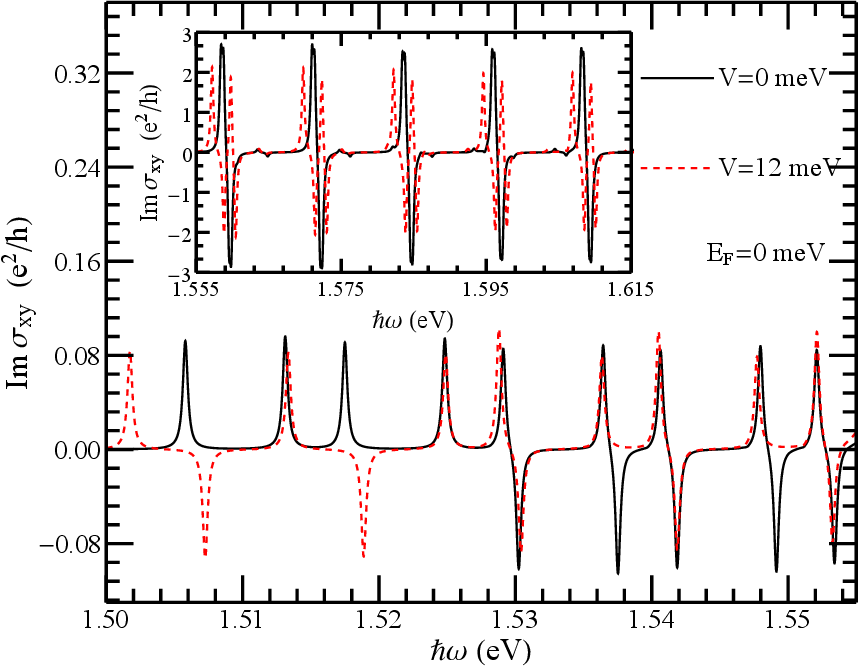}


\caption{As in Fig. 5 but for Im$\sigma_{xy}$. }
\label{f8}
\end{figure}

The electron energies are different than those of the holes  due to $\Delta$, the different values of the SOI and interlayer hopping (see Fig. 1). 
The terms intra-band and inter-band transitions refer to the bands in the absence of the magnetic field ($B = 0$).
In bilayer WSe$_{2}$ they belong to totally different regimes because of $\hslash \omega_{c} << \Delta$: the intra-band transitions fall in the microwave-to-THz regime and the inter-band ones in the visible frequency range because of the large value of the  gap $\Delta$. 
Unlike bilayer graphene-like 2D systems, the asymmetry between the CB and VB  in the bilayer WSe$_{2}$ spectrum, due to the huge band gap and strong SOI, has important implications for the peaks seen in Re$\sigma_{xx}^{nd}(\omega)$ and Im$\sigma_{xy}^{nd}(\omega)$ as  functions of the photon energy $(\hslash \omega)$.  

The absorptive part of the longitudinal conductivity is shown in Fig. 5 for a temperature $T = 0$ K and a level broadening $\Gamma = 0.04\sqrt{B}$ meV. A larger magnetic field ($B = 30$ T) has been used for well-resolved LL separation. The black solid and red dashed curves are for $E_{z}=0$ and  $E_{z}\neq 0$, respectively. Here, we took  $E_{F}=0$ eV in the gap.  The optical selection rules allow $n$ to change by only $1$, see Eqs. (\ref{l4})-(\ref{l5}). In addition, one needs to go from occupied $(n)$ to unoccupied $(n^{\prime})$ states through the absorption of photons with transitions allowed only between same-spin states. For $E_{z} = 0$ and $E_{z}\neq 0$, the series of peaks occur at $\hslash \omega = -E_{n+1,-,\mu_{2}}^{s,\tau} + E_{n,+,\mu_{2}}^{s,\tau}$ and $\hslash \omega = -E_{n,-,\mu_{2}}^{s,\tau} + E_{n+1,+,\mu_{2}}^{s,\tau}$ for integer $n$. This series of peaks corresponds to the allowed inter-band transitions in the LL structure.  As we can see from Fig. 5, the peaks are split due to the lifting of the spin and valley degeneracies in the presence of $B$ and absence of $E_{z}$ in contrast to the $B=0$ case. The spin-up transitions $-n\rightarrow (n+1)$ in $K$ $(K^{\prime})$ and spin-down ones $n\rightarrow -(n+1)$ in $K$ $(K^{\prime})$ are suppressed as seen by the small peaks in Fig. 5. On the other hand, the large peaks correspond to the spin-down transitions $-n\rightarrow (n+1)$ in $K$ ($K^{\prime}$) and the spin-up ones $n\rightarrow -(n+1)$ in $K$ $(K^{\prime})$. 

When   the  electric field is applied, 
the splitting of the peaks increases and the peaks move to lower energies as well as to higher energies. 
 Further, the spin and valley responses switch their labels. The shifting of peaks to lower energies  signals the reduction of  the band gap between CB and VB as can be seen in Figs. 1 and 2. Moreover, the shifting of the peaks to higher energies  signals an increase of the gap between  the  $E_{n,++}^{s,\tau}  ( E_{n,--}^{s,\tau})$ and   $E_{n,+-}^{s,\tau}  ( E_{n,-+}^{s,\tau})$  
 bands (see Figs. 1 and 2 ). As the electric field is turned on, the intensity of the peaks is reduced due to a redistribution of the spectral weight between the peaks as shown by the red dotted curve in Fig. 5. In contrast to monolayer WSe$_{2}$ \cite{rr13}, $\sigma_{xx}$ doesn't show any beating pattern at higher photon energies (not shown here) due to the well separated spin-up and spin-down states which do not mix at these frequencies. Another noteworthy point is that peak features in bilayer WSe$_{2}$ are completely different than in bilayer graphene \cite{rr45} due to the lack of perfect symmetry between the positive and negative branches (see Fig. 1). 
\begin{figure}[t]
\centering
\hspace{0.5cm}
\includegraphics[width=.45\textwidth]{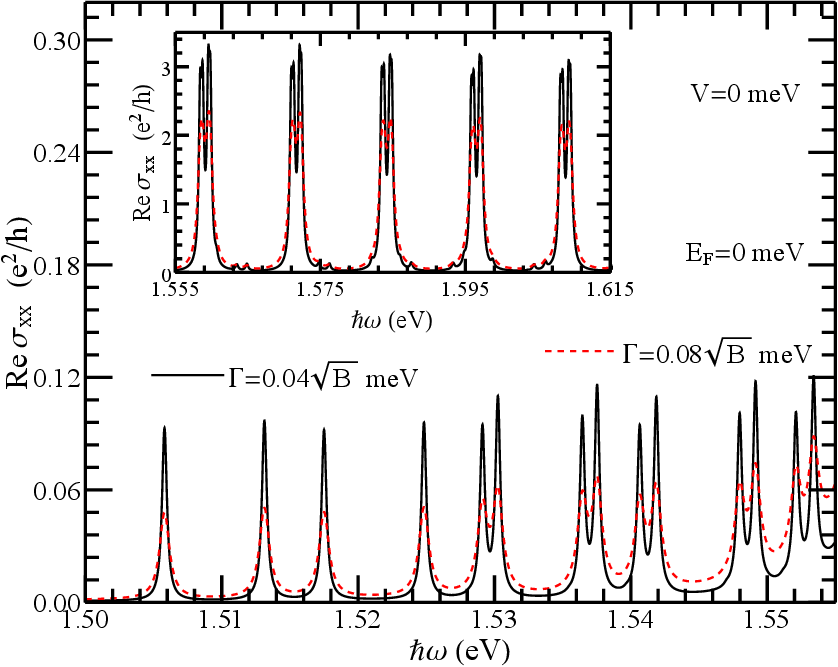}
\ \\
\ \\
\includegraphics[width=.45\textwidth]{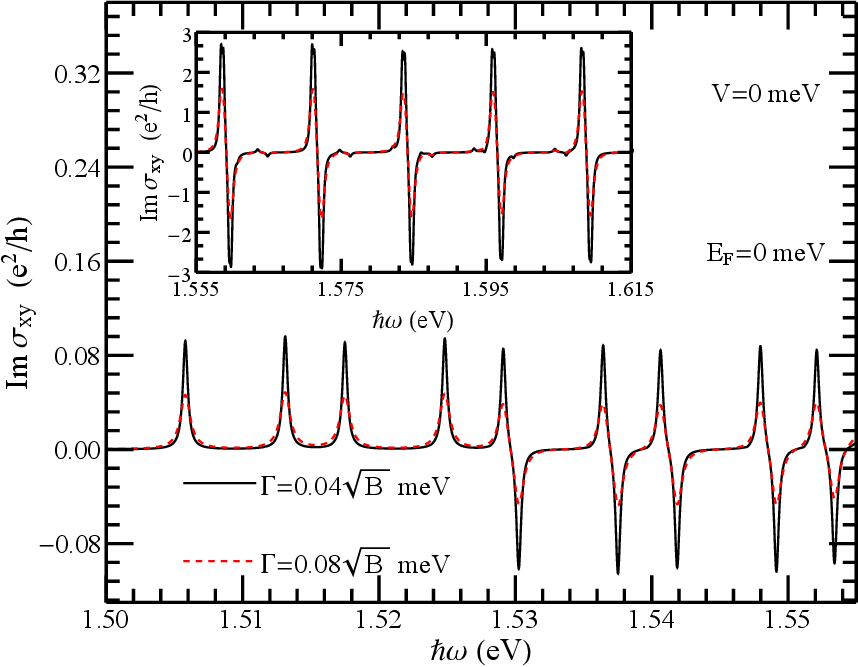}


\caption{The real part of $\sigma_{xx}$ (upper panel) and  the maginary part of $\sigma_{xy}$ (lower panel) 
vs $\hslash \omega$ for $V=0,\ E_{F}=0$ meV, $B=30$ T, and two values of the level width $\Gamma$.}
\label{f13}
\end{figure}

  A magnetic and electric control of the valley polarization can be clearly seen as the corresponding peaks in two different valleys appear at different frequencies. In addition to the valley-controlled transport, the peaks in each valley split as a result of all LLs becoming spin split. The spin and valley splittings can be understood with the help of Eq. (\ref{16}) and the corresponding energies. One noteworthy feature, that becomes clear by comparing the black and red curves of Fig. 5, is that the peaks are well separated for $E_{z}\neq 0$ in both  spin and valley spaces.  In  massless Dirac systems \cite{rr46}, the spin and valley peaks occur at the same frequency and hence a series of four peaks 
   is replaced by one peak  in contrast to bilayer WSe$_{2}$ shown in Fig. 5. It is obvious from Fig. 5 that real absorptive part of $\sigma_{xx}$ of the bilayer WSe has a much richer structure than its monolayer counterpart \cite{rr13}. 

The effect of  varying $E_{F}$ is shown in Fig. 6 for $E_{z}=0$. The  value $E_{F}=0.8358$ eV is situated between the $n=0$ and $n=1$ LLs, the first four peaks occurring at  $\hslash \omega< 1.53$ eV are completely removed due to Pauli blocking while all others ($\hslash \omega > 1.53$ eV) occur at the same energies as in  Fig. 5. This behaviour is opposite to that of other 2D materials \cite{rr46, rr47, rr48, rr49} like graphene, silicene, $\alpha-T_{3}$ and topological insulators, in  which the spectral weight of the inter-band peaks is continuously redistributed into the intra-band ones. This shows how the conductivity changes as $E_{F}$ moves through the LLs. Further, for $E_{z}\neq 0$ the lower peaks also disappear as $E_{F}$ moves to higher LLs.
\begin{figure}[t]
\centering

\includegraphics[width=.45\textwidth]{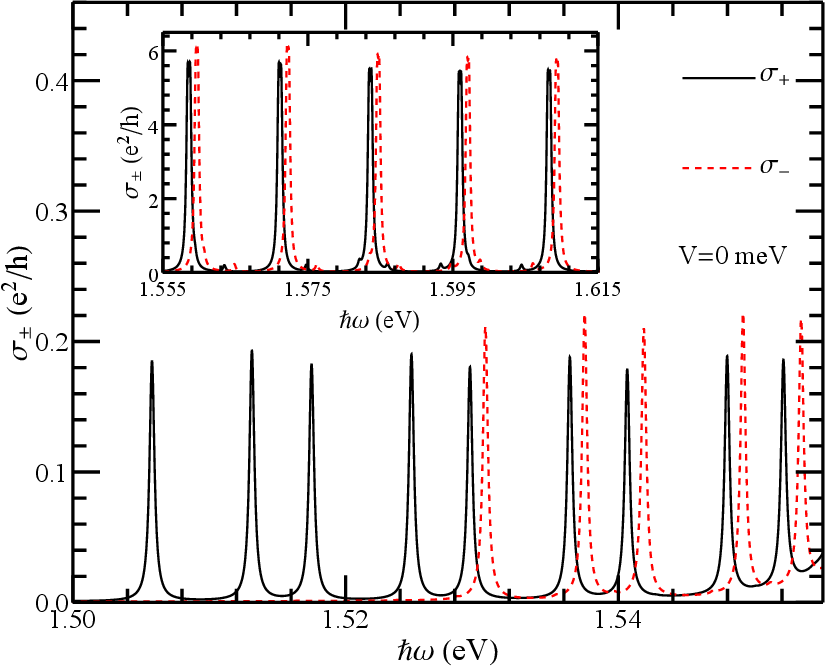}
\ \\
\ \\
\includegraphics[width=.45\textwidth]{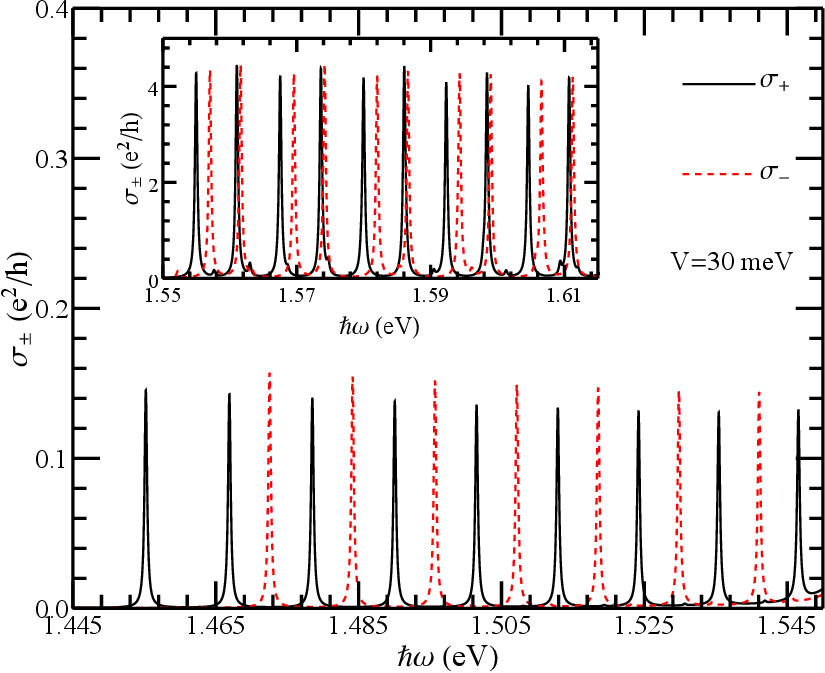}


\caption{Real part of the right polarized optical conductivity $\sigma_{+}(\omega)$ and of the left polarized one $\sigma_{-}(\omega)$ vs $\hslash \omega$ for $E_{F}=0$ meV and $B=30$ T. The solid black curve and red dotted one  are for $V=0$ meV and $V=30$ meV, respectively. }
\label{f1}
\end{figure}

For simplicity, we show   
 a $(B, \omega)$ contour plot of Re$\sigma_{xx}$ only 
for the $K$ valley in Fig. 7 aversus $B$  for two values of $V$: $V=0$ meV (upper panel) and  $V=30$ meV (lower panel). 
In bilayer WSe$_{2}$, as might be expected from Eq. (\ref{e4}), all observed transition energies behave  linearly with the  magnetic field $(\propto\hslash^{2}\omega_{c}^{2})$. In contrast, in bilayer graphene \cite{rr45} this occurs only for weak $B$  fields, but it switches over to a $\sqrt{B}$ dependence as the corresponding energy goes out of the parabolic band region. Also, the slope of the transition energies depends on the LL index $n$. In weak fields, the peaks are smeared out more easily in bilayer WSe$_{2}$ than in  its monolayer counterpart \cite{rr13}.  As expected, for $V \neq 0$, the peaks move to  lower values of $\hslash \omega$ due to the reduction of the gap between the CB and VB (cf. lower panel of Fig. 7).  

Figure 8 gives results for the Im$\ \sigma_{xy}$ as a function of energy $\hslash \omega$ in eV. The symmetry between positive and negative branches is no longer observed due to the $\Delta$ and SOI terms in Eq. (\ref{e1}), and the peaks corresponding to the transitions $-n\rightarrow (n+1)$ and $n\rightarrow -(n+1)$ have slightly different energies. Also, we can see the splitting of the  conductivity peaks due to these transitions. 
The strength of the splitting directly reflects  the energy difference 
between the CB and VB branches for the same $n$. The consequences of this difference 
 is even more striking for the Hall conductivity than it is for the longitudinal one. So, we can see this mismatch as emergence of positive and negative oscillations in conductivity. This behaviour can also be understood by the negative sign between the two terms of Eq. (\ref{16}). For the massless Dirac case, the negative and positive peaks would have the same energy and hence cancel out perfectly. Furthermore, there are no downward peaks 
 in the range $\hslash \omega<1.53$ eV for $E_{z}=0$  but 
 there are 
 when the field  $E_{z}$ is present.

In Fig. 9 we show the dependence of  Re$ \sigma_{xx}$ and  Im$  \sigma_{xy}$ on the values of $\Gamma$. The solid black curve is for broadening $\Gamma=0.04\sqrt{B}$ meV and the red dotted one for $\Gamma=0.08\sqrt{B}$ meV. The separation of the split peaks becomes narrow with increasing broadening $\Gamma$. By further increasing $\Gamma$, the splitting of the peaks disappears because the broadening covers the spacing between the spin-split LLs. To retain these peaks one has to apply a magnetic field for which the spin splitting exceeds the LL broadening $\Gamma \propto \sqrt{B}$. In other words, a large 
$\Gamma$ 
smears out the peaks.

The peak structure just described above for Re$\ \sigma_{xx}^{nd}(\omega)$ and Im$\ \sigma _{xy}^{nd}(\omega)$ importantly affects the behaviour of the conductivity for 
right $(+)$ and left $(-)$ polarized light. For real experiments that probe the circular polarization of resonant light, as in the case of the Kerr and Faraday effects, one evaluates the quantity $\sigma_{\pm}(\omega)$ given by 
\begin{equation}
\sigma_{\pm}(\omega) = \mathrm{Re}\sigma_{xx}^{nd}(\omega) \pm \mathrm{Im}\sigma_{xy}^{nd}(\omega),
\end{equation}
with the $+(-)$  sign corresponding to the right (left) polarization. In Fig. 10 we show $\sigma_{-}(\omega)$ (dotted red curve) and $\sigma_{+}(\omega)$ (solid black curve) as functions of the frequency,  for $E_{F }= 0.0$ eV in the gap, with $E_{z}=0$ (upper panel) and $E_{z}\neq 0$ (lower panel), using the parameters of Fig. 5. As seen, there is a direct correspondence between these results and those of Figs. 5 and 8. The heights of the peaks for $E_{z}=0$  and $E_{z}\neq 0$ in $\sigma_{-}(\omega)$ are slightly higher than those in $\sigma_{+}(\omega)$. Also, note that there is a double split-peak structure rather than a four split-peak structure as in $\sigma_{xx}(\omega)$. The peaks of $\sigma_{-}(\omega)$ and $\sigma_{+}(\omega)$ are displaced in energy with respect to  each other. 
 Similar to the behaviour of Re$ \sigma_{xx}^{nd}(\omega)$ and Im$ \sigma_{xy}^{nd}(\omega)$, the spin and valley splittings  increase  with  $E_{z}$. 

The difference between $\sigma_{-}(\omega)$ and $\sigma_{+}(\omega)$ is also reflected in the power absorption spectrum given by 
\begin{figure}[t]
\centering

\includegraphics[width=.45\textwidth]{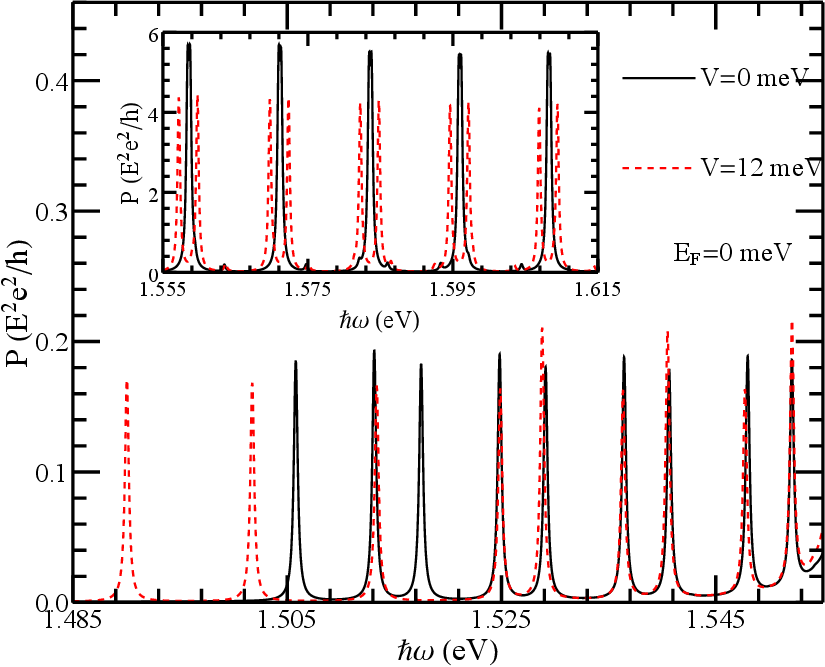}


\caption{Power spectrum vs $\hslash \omega$ for $V=0$ (black solid curve), $V=12$ meV (dotted red curve), and field $B=30$ T. }
\label{f11}
\end{figure}
\begin{equation}
P(\omega) = (E/2)\big[\sigma_{xx}(\omega) + \sigma_{yy}(\omega) − i\sigma_{yx}(\omega) + i\sigma_{xy}(\omega)\big].
\end{equation}
We recall that $\sigma_{\mu \nu} = \sigma_{\mu \nu}^{d }+ \sigma_{\mu\nu}^{nd} = \sigma_{\mu \nu}^{nd}$ since the component $\sigma^{d}_{\mu \mu}$, $\mu = x,y$, vanishes. The component $\sigma_{yy}^{nd}(\omega)$ is given by $\sigma_{xx}^{nd}(\omega)$ and $\mathrm{Im}\sigma_{xy}^{nd}(\omega) = '\mathrm{Im} \sigma_{yx}^{nd}(\omega)$. The spectrum $P(\omega)$ is shown in Fig. 11 as a function of the photon frequency for  $E_{z}=0$ and $E_{z}\neq 0$. Given that Im$ \ \sigma^{nd}_{ xy} (\omega)$ is  the negative of Re$\ \sigma^{nd}_{xx} (\omega)$, see Eq. (\ref{16}), the peaks in it are essentially the same as those in the longitudinal optical conductivity but   positive and negative. 
Similar to  Re$ \sigma^{nd}_{xx} (\omega)$ and Im$ \sigma^{nd}_{ xy} (\omega)$, spin and valley splittings can be clearly seen in Fig. 11 and
for $E_{z}\neq 0$ the separation between them increases.  

The semiclassical limit of the magneto-optical conductivity occurs when the magnetic field is very weak and the  spacing becomes inconsequential. This occurs for a large Fermi energy, $E_{F}\gg \varepsilon_{0,+-}^{s,\tau}$.  For $E_{F} >0$, only intra-band transitions are obtained between the $n$th and $(n + 1 )$th LLs in the CB. For $n\gg 0$, consider $E_{F} \approx E_{n,+,\mu_{2}}$ lies between the $n$th and $(n + 1)$th LLs. In this limit, the energy spacing is linear in $B$ in contrast to the $\sqrt{B}$ behaviour in Weyl semimetals \cite{, rr50}. The pertinent energy difference is $E_{n,+,\mu_{2}} - E_{n+1,+,\mu_{2}}=-\hslash \omega_{c}$.
\begin{figure}[t]
\centering

\includegraphics[height=3.5cm, width=6cm]
{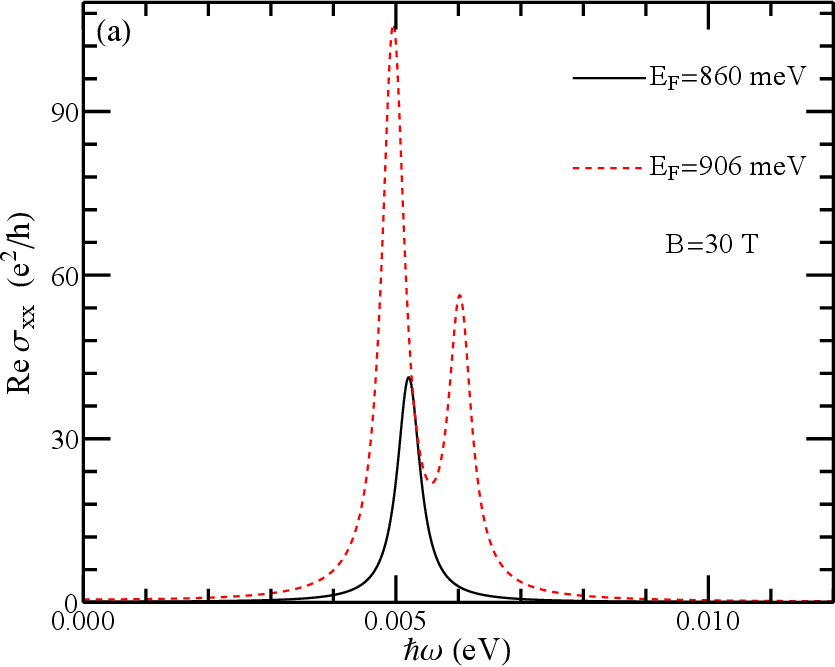}
\ \\
\ \\
\includegraphics[height=3.5cm, width=6cm]
{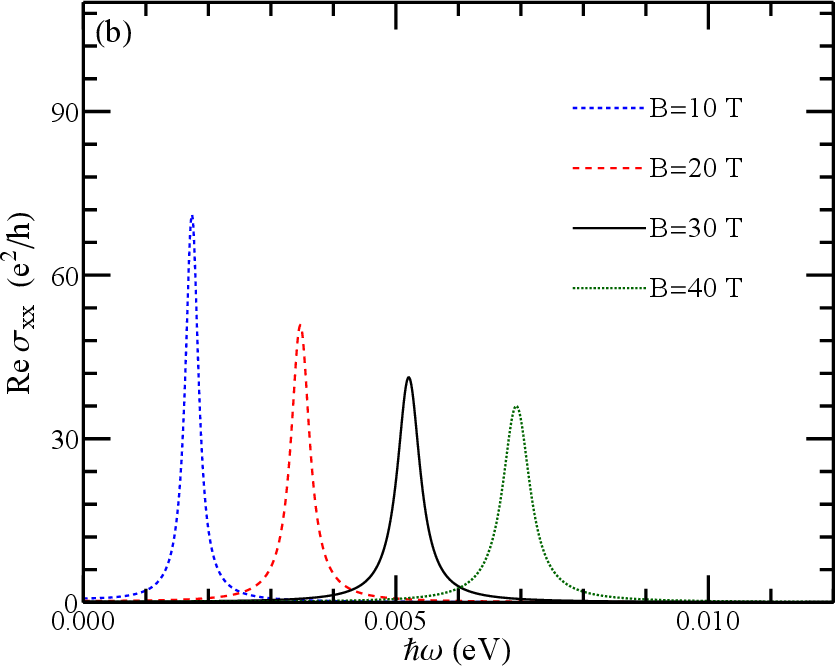}


\caption{(a) Intra-band limit of the real part of the longitudinal 
conductivity versus photon energy $\hslash \omega$ for  $B=30$ T and two values of $E_{F}$. (b) As in (a) for four values of $B$  and $E_{F}$ close to $860$ meV for $B\neq 30$ T.  The energy $\hslash \omega$ is measured from the bottom of the conduction band. }
\label{f14}
\end{figure}

We show the results of Re$ \sigma_{xx}^{nd}(\omega)$ for the intra-band case in Fig. 12. We  see from the upper panel  that there is a spectral weight redistribution to a strong intra-band response when $E_{F}$ increased. Furthermore, the optical spectral weight under these curves increases with $E_{F}$ in contrast to topological insulators \cite{rr46}.  Further, a double peak response is present in the strong intra-band response as  the dashed red curve in the upper panel shows. This results from the spin splitting of the LLs that renders 
the spin levels at a given valley unequal in energy. 
 Also, the separation between the double peaks (red dashed curve) increases with 
 $E_{F}$. 
Similar to the monolayer WSe$_{2}$ \cite{rr13}, these peaks lie in the range of microwave-to-THz frequencies and their height is larger than that of the inter-band transitions shown in Figs. 5-11.  Further, when we increase the magnetic field $B$, as  seen in the lower panel of Fig. 12, the intra-band peaks move to higher energies and their height is reduced in contrast to massless Dirac materials \cite{rr46, rr47}.  For   large $E_{F}$  the effect of $E_{z}$ becomes inconsequential. These results are consistent with graphene-like 2D systems in which the relevant spectral weight increases with $E_{F}$, while the optical features in these 2D systems lie only in the THz regime  \cite{rr46, rr47, rr48, rr49, rr50}.

\section{summary and conclusions}

We have shown how the gap $\Delta$ and the SOI strength modify the electronic energy dispersion in bilayer WSe$_{2}$ , unlike bilayer graphene \cite{rr41, rr42}, in the absence and presence of magnetic and electric fields. For $B=E_{z}=0$ and $B\neq E_{z}\neq 0$, the energies of the levels in the conduction and valence bands no longer mirror each other, cf. Figs. 1, 2. Further, we have studied the spin- and valley-controlled magnetotransport in the presence and absence of $E_{z}$. We point out that inter-band optical transitions from level $n$ in the valence band to level $n + 1$ in the conduction band no longer have the same energy as those from level $n + 1$ to  level $n$;  this splits the corresponding absorption line in the real part of the longitudinal conductivity. Also, the optical spectral weight of these lines is different (see the large and small peaks of Fig. 4) from that in graphene. 
The energy of the splitting is related to the mismatch in energy levels between the conduction and valence bands, see Fig. 2. A similar splitting was found for the imaginary part of the Hall conductivity. 

Due to the large $\Delta$, $\lambda_{c}$ and $\lambda_{v}$ terms, the conductivity peaks in WSe$_{2}$ depend linearly on $B$, contrary to bilayer graphene \cite{rr45}, and reflect the equidistant LLs in each band. 
In addition, the onset energies of the spin- and valley-dependent transitions  reflect 
the energy difference between the LLs and are controlled by the magnetic and electric fields. The other determining factors are the band gap and the SOI strength. Accordingly, we may expect that a careful tuning of electric and magnetic fields will 
determine the value of band gap and SOI strength.  However, for the absorption of circularly polarized light, two-peak structures are recovered but in this case there is a shift in the energy position and amplitude of the lines between right and left polarizations in contrast to what is found when the band gap and SOI terms in the electron dispersion curves are zero for graphene. 

The semiclassical limit is affected by the 
magnetic field. This significantly shifts not only the intra-band  peak to higher $\hslash \omega$ values, but also reduces the peak am-\\plitude   in contrast with graphene. The lineshape associ-\\ated with the intra-band magneto-conductivity is significantly changed when the Fermi energy is varied. The optical spectral weight under these curves is found to increase in contrast to topological insulators and similar 
massless 
 Dirac systems \cite{rr46}. These novel findings  may be pertinent to the development of 
 spintronic and valleytronic optical devices based on  bilayer TMDCs. \\

\acknowledgments
M. Z. and P. V. acknowledge the support of the Canadian NSERC Grant No. OGP0121756. The work of M. T. was supported by  Colorado State University.

\appendix
\begin{widetext}

\section{Zero-level Hall conductivity}

Using Eq. (\ref{e6}), and Eq. (\ref{e8}) the off-diagonal velocity matrix elements for  $n=0$ are
\begin{eqnarray}
\left\langle 0,\mu,s,\tau\right\vert v_{x}\left\vert n^{\prime},\mu^{\prime
},s^{\prime},\tau^{\prime}\right\rangle   =\tau v_{F }\ \varrho_{0,\mu}^{s,\tau
}\varrho_{n^{\prime},\mu^{\prime}}^{s^{\prime}, \tau^{\prime}} \delta_{s,s^{\prime}}
  \times 
\Big\{
\sqrt{n^{\prime}}/\varepsilon_{n,d_{2}^{\prime} }  +
k_{0,\mu}^{s,\tau} k_{n^{\prime},\mu^{\prime}}^{s^{\prime},\tau^{\prime}} /\varepsilon_{0,d_{4}}    \Big\} \ \delta_{0,n^{\prime}-1}   
\end{eqnarray}
\begin{eqnarray}
\left\langle n^{\prime},\mu^{\prime},s^{\prime},\tau^{\prime}\right\vert
v_{y}\left\vert 0,\mu,s,\tau\right\rangle    =\tau i v_{F }\ \varrho_{0,\mu}^{s,\tau
}\varrho_{n^{\prime},\mu^{\prime}}^{s^{\prime}, \tau^{\prime}} \delta_{s,s^{\prime}}
  \times
\Big\{
\sqrt{n^{\prime}}/\varepsilon_{n,d_{2}^{\prime} }   +
k_{0,\mu}^{s,\tau} k_{n^{\prime},\mu^{\prime}}^{s^{\prime},\tau^{\prime}} /\varepsilon_{0,d_{4}}    \Big\} \ \delta_{0,n^{\prime}-1}   
\end{eqnarray}
\begin{equation}
\left\langle 0,+-,\tau \right\vert v_{x}\left\vert n^{\prime},\mu^{\prime
},s^{\prime},\tau^{\prime}\right\rangle =\tau v_{F}Y, 
\quad
\left\langle n^{\prime},\mu^{\prime},s^{\prime},\tau^{\prime}\right\vert
v_{y}\left\vert 0 ,+-,\tau \right\rangle =\tau iv_{F}Y, \quad Y=\varrho_{n^{\prime},\mu^{\prime}%
}^{s^{\prime}, \tau^{\prime}} k_{n^{\prime},\mu^{\prime}}^{s^{\prime},\tau^{\prime}}\delta_{s,s^{\prime}}\,\delta_{0,n^{\prime}},
\end{equation}
Using these expressions the 
conductivities take the form
\begin{eqnarray}
\begin{pmatrix}
\mathrm{Re}\ \sigma_{xx}^{nd}\\
\ \\
\mathrm{Im}\ \sigma_{xy}^{nd}
\end{pmatrix}    =\mp \frac{e^{2}}{2h}
\sum_{s,\tau,\mu,\mu^{\prime}}  \eta_{0,1,\mu,\mu^{\prime}}^{s,\tau}  
\Big[ \dfrac{1}{\bigl(\varepsilon_{0,\mu}^{s, \tau}-\varepsilon_{1,\mu}^{s, \tau}+\bar{\omega} \bigr)^{2}+\bar{\Gamma}^{2}} \pm \dfrac{1}{\bigl(\varepsilon_{0,\mu}^{s, \tau}-\varepsilon_{1,\mu}^{s, \tau}-\bar{\omega} \bigr)^{2}+\bar{\Gamma}^{2}}  \Big],\notag
\\*
%
\, \, \, =\mp \frac{e^{2}}{2h}
\sum_{s,\tau,\mu^{\prime}}   \upsilon_{0,+-,\mu^{\prime}}^{s,\tau}  \Big[ \dfrac{1}{\bigl(\varepsilon_{0,+-}^{s, \tau}-\varepsilon_{0,\mu}^{s, \tau}+\bar{\omega} \bigr)^{2}+\bar{\Gamma}^{2}} \pm \dfrac{1}{\bigl(\varepsilon_{0,+-}^{s, \tau}-\varepsilon_{0,\mu}^{s, \tau}-\bar{\omega} \bigr)^{2}+\bar{\Gamma}^{2}}  \Big] ,\label{A1}
\end{eqnarray}
where
\begin{equation}
\label{A2}
\eta_{0,1,\mu,\mu^{\prime}}^{s, \tau}=  \bar{\Gamma}\bigl(  \varrho_{0,\mu}^{s, \tau}\varrho_{1,\mu^{\prime}}^{s, \tau
}\bigr)  ^{2}  \Big[  \frac{1}{\varepsilon_{1,d_{2} }^{\prime} }   +\frac{k_{0,\mu}^{s,\tau} k_{1,\mu^{\prime}}^{s,\tau} } {\varepsilon_{0,d_{4}} }   \Big]^{2}\frac{
f_{0,\mu}^{s, \tau}-f_{1,\mu^{\prime}}^{s,\tau
}
 }{
 \varepsilon_{0,\mu}^{s, \tau}-\varepsilon_{1,\mu^{\prime}}^{s, \tau
}
},
%
\quad\quad\upsilon_{0,+-,\mu^{\prime}}^{s,\tau}=\bar{\Gamma} \bigl(  \varrho_{0,\mu^{\prime}}
^{s, \tau} k_{0,\mu^{\prime}}^{s,\tau} \bigr)  ^{2}\ \frac{
f_{0,+-}^{s, \tau
}-f_{0,\mu^{\prime}}^{s, \tau}
 }{
\varepsilon_{+-}^{s,\tau}-\varepsilon_{0,\mu^{\prime}}^{s, \tau}
}
\end{equation}

\end{widetext}

\end{document}